\documentclass{article}
\usepackage{spconf,amsmath,epsfig}

\usepackage[]{graphicx}
\usepackage{caption}
\usepackage{subfig}
\usepackage{amsmath}
\usepackage{amssymb}
\usepackage{epsfig}
\usepackage{cite}
\usepackage{color}
\usepackage{balance}
\usepackage{amsmath}
\usepackage{amsfonts} 
\usepackage{graphicx}
\usepackage{pdfpages}
\usepackage[T1]{fontenc} 
\usepackage{array}
\usepackage{url}

\usepackage{color}
\usepackage{wrapfig}
\usepackage{lipsum}
\usepackage[hidelinks]{hyperref}
\usepackage{multirow}
\usepackage{tabularx}
\usepackage{gensymb} 
\usepackage{breqn}
\usepackage{array}
\usepackage{booktabs} 
\usepackage{textcomp}

\title{Systematic Study of Weather Variables for Rainfall Detection}

\name{Shilpa Manandhar$^{\dagger}$$^{1}$, Soumyabrata Dev$^{\dagger}$$^{2}$, Yee Hui Lee$^{1}$, Stefan Winkler$^{3}$ and Yu Song Meng$^{4}$
\thanks{$^{\dagger}$~Authors contributed equally.}
\thanks{This research is funded by the Defence Science and Technology Agency (DSTA), Singapore.}
\thanks{Send correspondence to Y.\ H.\ Lee, E-mail: EYHLee@ntu.edu.sg.}
}
\address{
	$^{1}$~School of Electrical and Electronic Engineering, Nanyang Technological University, Singapore \\
	$^{2}$~ADAPT Centre, School of Computer Science and Statistics, Trinity College Dublin, Ireland \\
    $^{3}$~Advanced Digital Sciences Center (ADSC), University of Illinois at Urbana-Champaign, Singapore\\
    $^{4}$~National Metrology Centre, Agency for Science, Technology and Research (A$^{*}$STAR), Singapore \\
}

\begin{document}

\maketitle

\begin{abstract}
Numerous weather parameters affect the occurrence and amount of rainfall. Therefore, it is important to study these parameters and their interdependency. In this paper, different weather and time-related variables -- relative humidity, solar radiation, temperature, dew point, day-of-year, and time-of-day are analyzed systematically using Principal Component Analysis (PCA). We found that four principal components explain a cumulative variance of 85\%. The first two principal components are applied to distinguish rain and no-rain scenarios as well. We conclude that all 7 variables have similar contribution towards rainfall detection. 
\end{abstract}

\begin{keywords}
Precipitation, weather sensor, remote sensing, PWV
\end{keywords}

\section{Introduction}
\label{sec:intro}
Precipitation is influenced by different factors to varying degree. Recently, because of the growing proliferation of geostationary satellites, there has been renewed interest in using GPS signal delays to better understand this phenomenon. Yao et al.\ \cite{Yibin} and Benevides et al.\ \cite{Benevides} studied GPS delays to predict rainfall. In our recent paper \cite{GPS_shilpa}, we have also discussed the importance of GPS-derived precipitable water vapor (PWV) for understanding rainfall. 

However, the classification of rain and no-rain scenarios using only GPS PWV values may not result in good accuracy. In \cite{IGARSS_Shilpa}, we have illustrated the usefulness of different weather variables in rainfall detection. For example, during rainy conditions, temperature generally approaches the dew point; relative humidity reaches nearly 100\%, and  precipitable water vapor increases. Solar radiation usually decreases prior to rain and during rain because of the presence of clouds.

In this paper, we first provide a systematic analysis of various weather parameters and use them in detecting precipitation. The source code of all simulations is available online.\footnote{~\url{https://github.com/Soumyabrata/weather-features}}

\section{Weather Variables}
\label{sec:parameters}

\subsection{Surface Weather Parameters}
Surface weather parameters recorded by a weather station (Davis Instruments 7440 Weather Vantage Pro II) with tipping bucket rain gauge are used in this study. The weather station is located at Nanyang Technological University (NTU), (1.3$^{\circ}$N, 103.68$^{\circ}$E). The weather station measurements used for this paper include Surface Temperature $^{\circ}C$, Relative Humidity ($RH$) (\%), Dew Point $^{\circ}C$, Solar Irradiance ($W/m^{2}$), and Rainfall Rate (mm/hr). All weather parameters are recorded at 1-minute intervals. 

\subsection{GPS Derived Water Vapor Content}
In addition to the various surface weather parameters, precipitable water vapor (PWV) values derived from GPS signal delays are used as an additional important feature for classification. This section provides a brief overview of the derivation of PWV values from GPS signal delays. 

PWV values (in mm) are calculated using the zenith wet delay (ZWD), \textit{$\delta$L$_w^{o}$}, incurred by the GPS signals as follows: 
\begin{equation}
	\mbox{PWV}=\frac{PI \cdot \delta L_w^{o}}{\rho_l},
    \label{eq1}
\end{equation}     
where $\rho_{l}$ is the density of liquid water (1000 kg$/m^{3}$), and \textit{PI} is the dimensionless factor determined by \cite{shilpaPI}:
\begin{dmath}
	PI=[-\text{sgn}(L_{a})\cdot 1.7\cdot 10^{-5} |L_{a}|^{h_{fac}}-0.0001]\cdot \\ 
    \cos \frac{2\pi(DoY-28)}{365.25}+
    0.165-1.7\cdot 10^{-5}|L_{a}|^{1.65}+f,
    \label{eq2}
\end{dmath}
where $L_a$ is the latitude, \textit{DoY} is day-of-year, $h_{fac}$ is either 1.48 or 1.25 for stations in the Northern or Southern hemisphere, respectively.  $f=-2.38\cdot 10^{-6}H$, where $H$ is the station altitude above sea level, can be ignored for $H<1000m$. 

For this paper, the ZWD values for an IGS GPS station located at NTU (station ID: NTUS) are processed using GIPSY OASIS software and recommended scripts \cite{GIPSY}. PWV values are then calculated for NTUS using Eqs.~\ref{eq1}-\ref{eq2}, with $L_a= 1.34$, $h_{fac}= 1.48$, $H=78m$. The calculated PWV values have a temporal resolution of 5 minutes. 

\subsection{Seasonal and Diurnal Features}
Singapore has tropical weather conditions, with scattered to abundant rainfall throughout the year. It has two distinct monsoon seasons, and witnesses uniform temperature throughout the year. Since seasons play an important role in rainfall, and rainfall occurrences in tropical regions also show clear diurnal characteristics. Thus, in addition to the weather sensor and GPS PWV data, seasonal and diurnal variables are included in the analysis the form of Day-of-Year (\textit{DoY}) and Time-of-Day (\textit{ToD}).

In this paper, all of the above mentioned variables and the rain data for the year $2010$ are used as our experimental dataset. Since GPS PWV values have a resolution of $5$ minutes, all surface weather parameters are also sampled at 5-minute intervals.

\section{Interdependency of Variables}

\begin{figure}[htb]
\begin{center}
\includegraphics[width=0.45\textwidth]{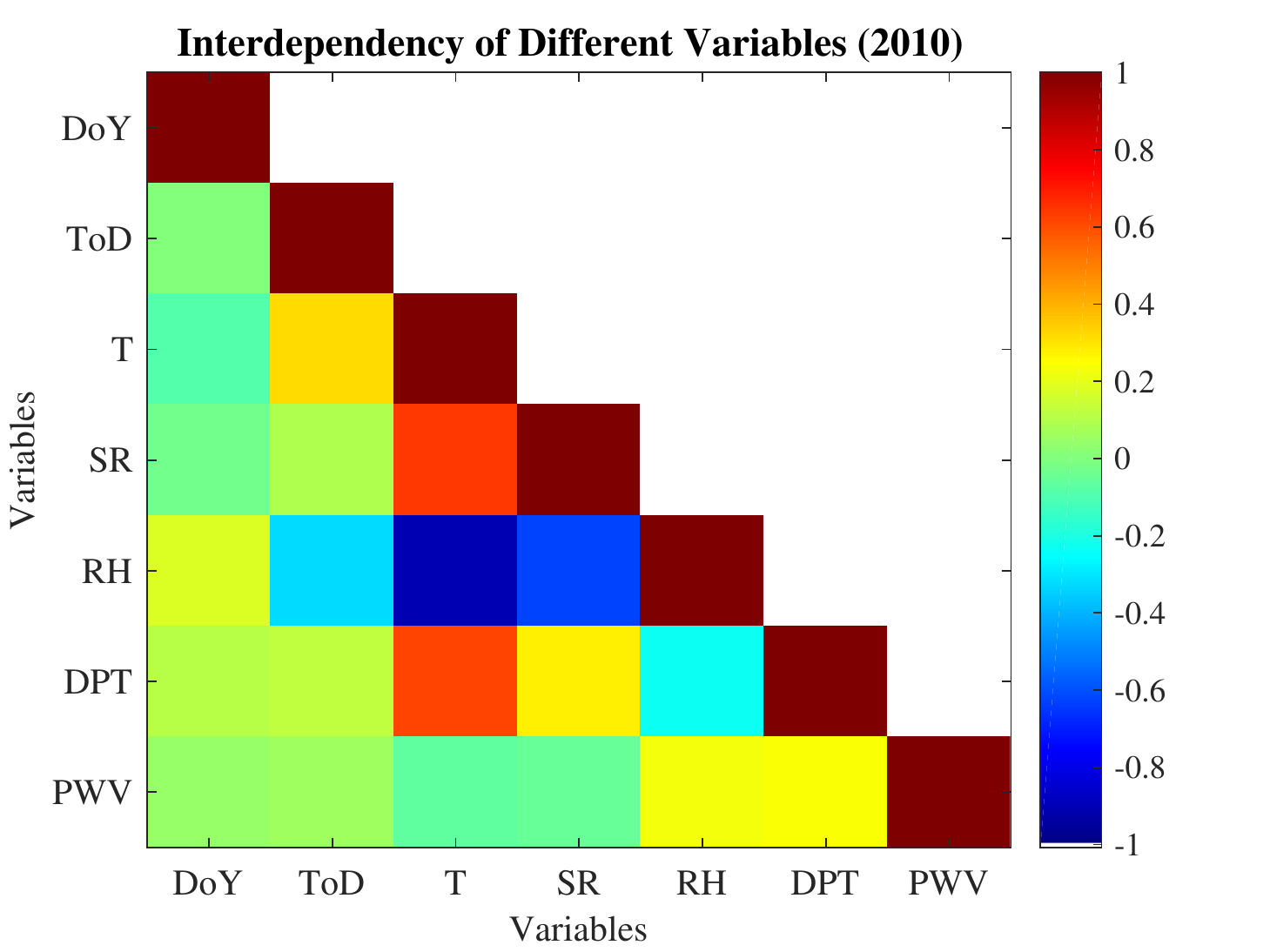}
\caption{Correlation between various features (best viewed in color).
\label{fig:cmat}}
\end{center}
\vspace{-0.8cm}
\end{figure}

Based on the database, a $7$-dimensional feature vector comprising relative humidity (RH), solar radiation (SR), temperature (T), dew point (DPT), precipitable water vapor (PWV), day of the year (\textit{DoY}) and time of the day (\textit{ToD}) is formulated. The degree of cross-correlation amongst these variables is calculated and shown in Fig.~\ref{fig:cmat}. It is important to analyze such correlation trends before any classification task is performed, because if two features are perfectly correlated, then the second feature does not contribute any additional information (it is already determined by the other). 

Various observations can be made from the off-diagonal elements. T has a good correlation with RH, SR, and DPT. The correlation between T and RH is the highest with a correlation coefficient of $-0.9$, which indicates that the two variables are highly negatively correlated. T $\&$ SR and T $\&$ DPT have positive correlation coefficients of $0.6$. RH and SR are negatively correlated with a correlation coefficient of $-0.6$. Solar radiation is lowest in the night, and relative humidity at night on a tropical island like Singapore is generally very high. Therefore a high negative correlation is observed between the two variables, dependent on \textit{ToD}.

It is also interesting to note that the PWV does not show strong correlation with any of the other variables. However, it has a small positive correlation with RH and DPT. Different seasons and locations have a direct impact on the behavior and correlation of these variables. For temperate regions, a higher degree of correlation is observed between RH and PWV\cite{RH_PWV}, whereas for tropical regions (viz.\ Singapore), the weather parameters exhibit a very different behavior.

\section{Principal Component Analysis}
\label{sec:pca-theory}
We use Principal Component Analysis (PCA) to investigate the underlying structure of the 7 weather variables. It is also used to determine the individual contributions of these variables to rainfall. 

Consider a variable matrix $\mathbf{X}$ of dimension $m \times n$, where $m$ is the number of variables (7 in this case) and $n$ is the number of observations. The individual variables $v_{1-7}$ are extracted from $\mathbf{X}$ and reshaped into the column vectors $\mathbf{\widetilde{v}}_j \in {\rm I\!R}^{mn \times 1}$ where $j=1,2,..,7$. The $\mathbf{\widetilde{v}}_j$ obtained from $\mathbf{X}$ are stacked alongside, to form the matrix  $\hat{\textbf{X}} \in {\rm I\!R}^{mn \times 7}$:
\begin{equation}
\label{eq:eq1}
\hat{\textbf{X}}=[\mathbf{\widetilde{v}}_1, \mathbf{\widetilde{v}}_2,\mathbf{\widetilde{v}}_3,\mathbf{\widetilde{v}}_4,\mathbf{\widetilde{v}}_5,\mathbf{\widetilde{v}}_6,\mathbf{\widetilde{v}}_{7}].
\end{equation}
$\hat{\textbf{X}}$ is then normalized with respect to the mean $\bar{v_{j}}$ and standard deviation $\sigma_{v_{j}}$ of the individual variables of $\mathbf{X}$, and is represented by $\ddot{\mathbf{X}}$:
\begin{equation}
\label{eq:eq3}
\ddot{\mathbf{X}}= \left[\frac{\widetilde{\mathbf{v}_{1}}-\bar{v_{1}}}{\sigma_{v_{1}}}, \frac{\widetilde{\mathbf{v}_{2}}-\bar{v_{2}}}{\sigma_{v_{2}}},..,\frac{\widetilde{\mathbf{v}_{j}}-\bar{v_{j}}}{\sigma_{v_{j}}},..,\frac{\widetilde{\mathbf{v}_{7}}-\bar{v_{7}}}{\sigma_{v_{7}}}\right].
\end{equation}

Subsequently, the covariance matrix of $\ddot{\mathbf{X}}$ is computed. The eigenvalue decomposition of the covariance matrix then yields the eigenvalues and the eigenvectors. The eigenvectors define the new orthogonal axes called principal components (PC). The eigenvalues explain the variance contributed by each of the $7$ principal components (PC 1-7).

\subsection{Variance Explained by Principal Components}
\begin{figure}[htb]
\begin{center}
\includegraphics[width=0.45\textwidth]{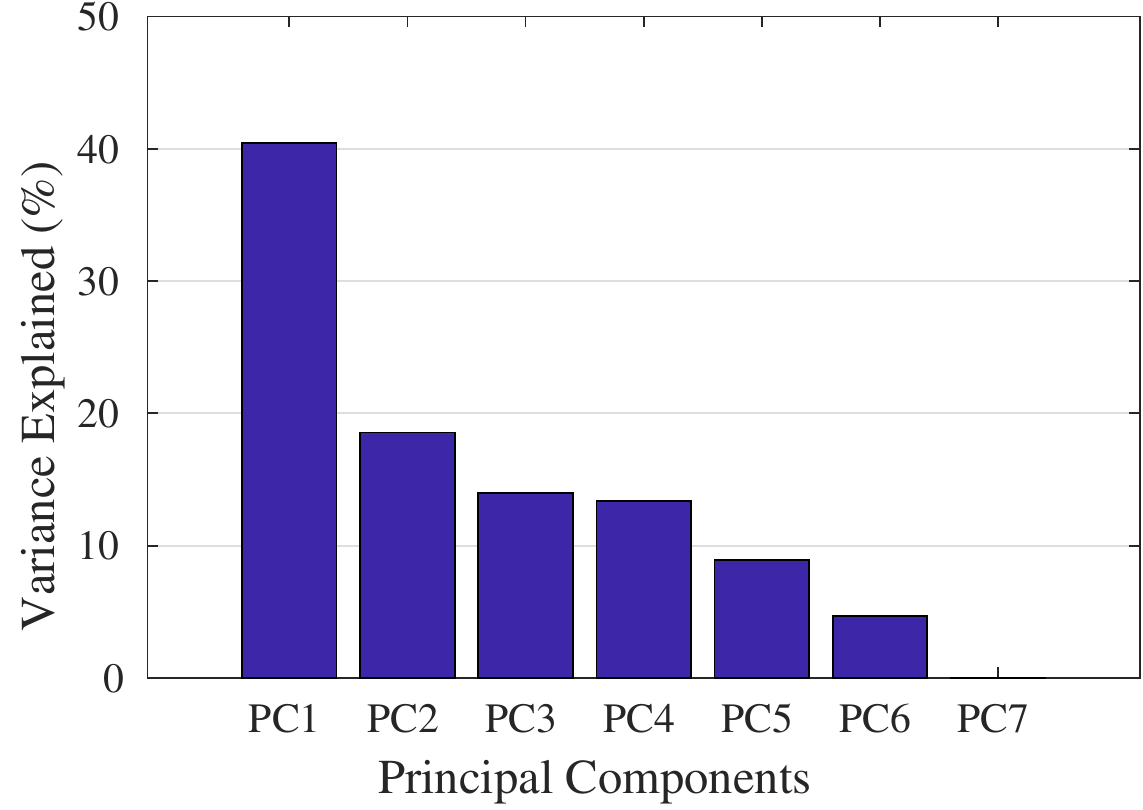}
\caption{Variance explained by different Principal Components.
\label{fig:variance_explained}}
\end{center}
\vspace{-1.1cm}
\end{figure}

In this section, the most important principal components are analyzed based on the variance explained by each. The percentage of total variance explained by each PC is calculated and shown in Fig.~\ref{fig:variance_explained}. About 40\% of the variance is explained by the first principal component (PC1). PC1-PC4 together explain a cumulative variance of 85\%. 

\subsection{Biplot Representation}

\begin{figure}[htb]
\begin{center}
\includegraphics[width=0.46\textwidth]{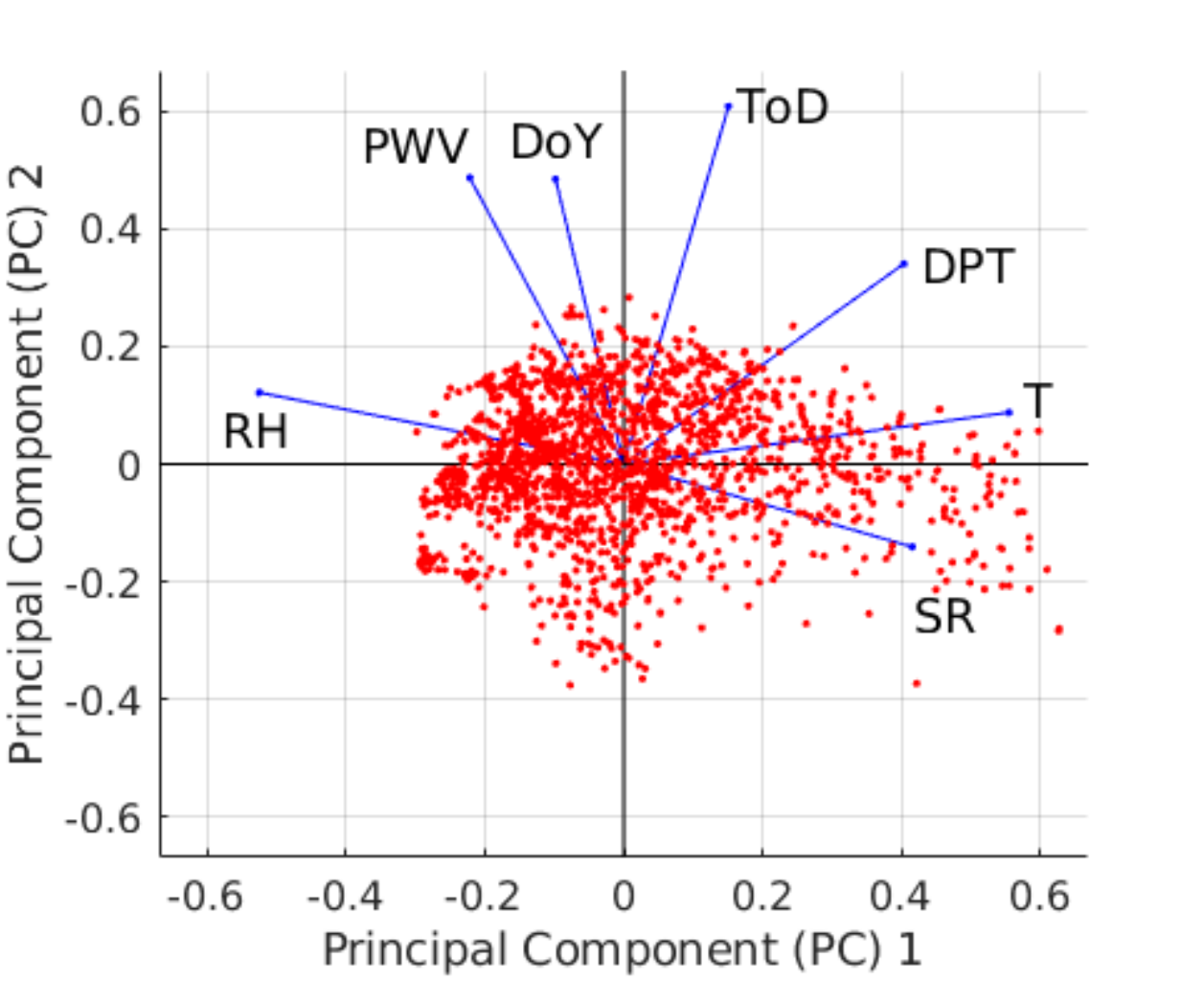}
\caption{Biplot representation of the weather variables. The red points indicate the different observations in our dataset, and the vectors represent the different weather variables.
\label{fig:biplot}}
\end{center}
\vspace{-1cm}
\end{figure}

Fig.~\ref{fig:biplot} visualizes the distribution of the samples in the new principal component axes via a biplot representation. This biplot provides us interesting insights on our weather variables, and their contribution to the principal components. It shows the contribution of each of the weather variables to the two principal components, and the correlation amongst them. As an illustration, PWV and $DoY$ are positively correlated with each other, whereas T and RH are negatively correlated to each other. The data points are represented as \emph{red} points in the transformed vector subspace, comprising the two principal components. 

\subsection{Variable Loadings}
Variable loadings are the contribution of different variables to an individual principal component. Table \ref{table:variable_loadings} lists the loading factors for the first four PCs, which together explain 85\% of the cumulative variance (see above). 

\begin{table}[htb]
\centering
\caption{Loading factors of the different variables for the first four principal components. Variables with above-average contribution are marked in bold.}
\begin{tabular}{*{5}{c}}
\toprule
\multicolumn{1}{c|}{Variables} & \multicolumn{1}{c|}{PC1} & \multicolumn{1}{c|}{PC2}& \multicolumn{1}{c|}{PC3}& \multicolumn{1}{c}{PC4}\\
\midrule
{T}  & \textbf{0.57}    & 0.03            & 0.03    & -0.04 \\
{RH} & \textbf{-0.53}   &0.24   & 0.009    & -0.05 \\
{SR} & \textbf{0.44} &-0.06  & 0.22     & -0.22\\
{DPT}& 0.33    &\textbf{0.54}    & 0.07      & -0.02\\
{ToD}& 0.23     & 0.08 & \textbf{-0.47} & \textbf{0.79}\\
{DoY}&-0.06 & \textbf{0.41} & \textbf{0.73} & \textbf{0.47}\\
{PWV}&-0.05&\textbf{0.68}&\textbf{-0.42}& -0.22\\
\hline
\end{tabular}
\label{table:variable_loadings}
\end{table}

The sum of the squares of all loadings for an individual principal component equals unity. A positive (negative) loading indicates that the variable and the particular principal component are positively (negatively) correlated. A large (either positive or negative) loading factor indicates that a variable has a strong effect on that principal component. The threshold for loading is calculated by considering equal contribution from all variables to the principal component. In our case, the number of variables is $7$, and therefore the loading threshold is $\sqrt{1/7}=0.378$. Any variable that has a loading above this threshold can be be considered an important contributor to that principal component. Those variables are highlighted in Table \ref{table:variable_loadings}.

All principal components are orthogonal to each other and hence uncorrelated. Therefore, each individual PC can be useful to explain a unique phenomenon. In our case, PC1 has strong positive loadings for T and SR and has a strong negative loading for RH. These variables depict changes with rainfall. When it rains, humidity goes higher, temperature and solar radiation both drop \cite{IGARSS_Shilpa}. Thus, PC1 might be useful in classifying rainfall scenarios. 

PC2 has strong positive loading for PWV, \textit{DoY} and DPT. It might be useful in explaining the seasonal dependency of PWV and its correlation to DPT. PWV values are generally higher in rainy seasons and lower in non-rainy seasons. Thus, PC2 might also explain the seasonal dependency of rainfall. 

The remaining principal components PC3 and PC4 have strong positive loadings for \textit{DoY} and \textit{ToD} respectively. Thus, these PCs are useful in explaining the temporal variations of different phenomenon. It can also be useful in explaining the nature of occurrence of frequent evening rain events in Singapore during the North-East Monsoon and early morning downpours during the Inter-Monsoon seasons.

\subsection{Rainfall Detection}
\begin{figure}[htb]
\begin{center}
\includegraphics[width=0.45\textwidth]{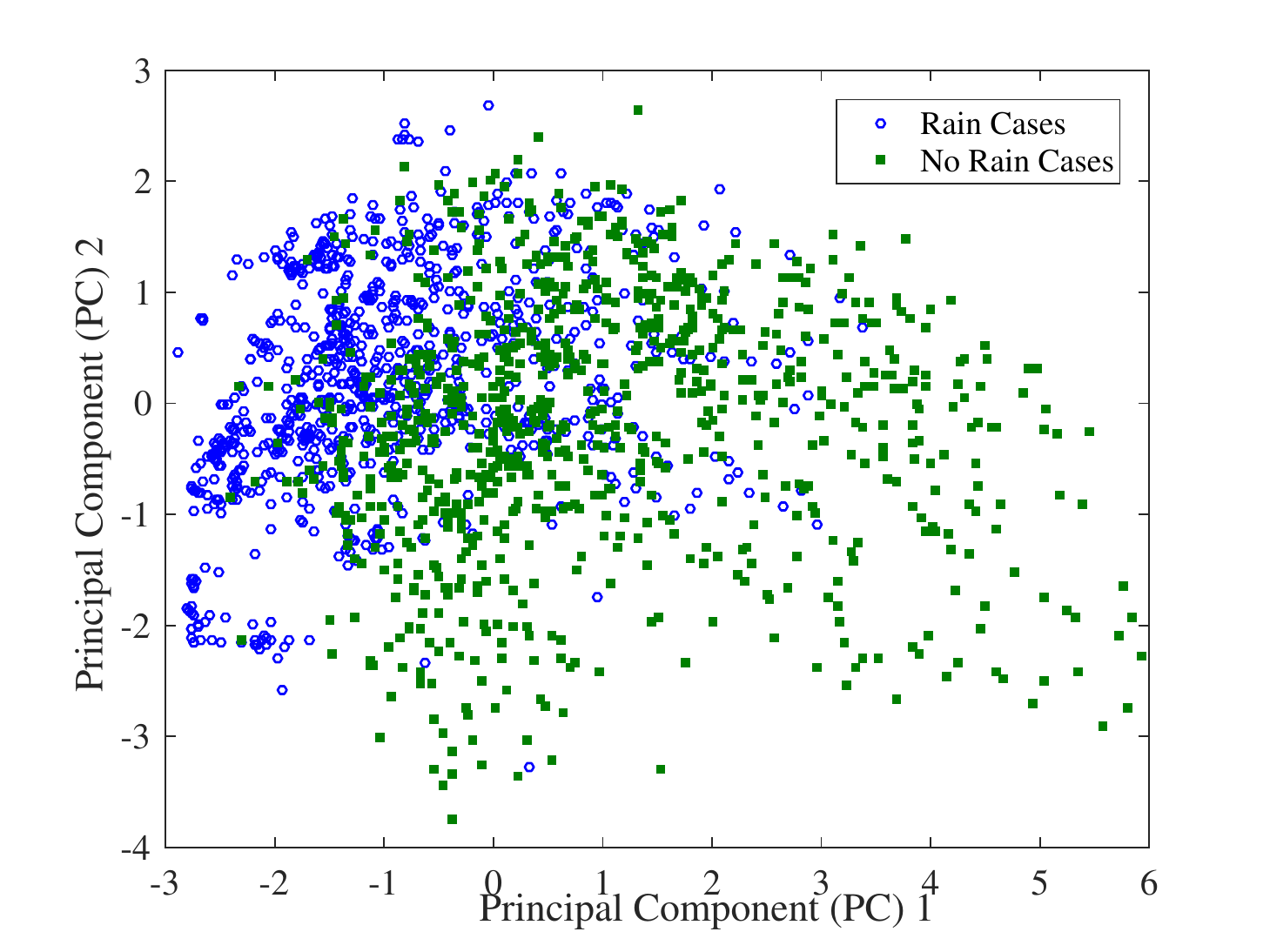}
\caption{Visualization of high-dimensional features in a lower dimensional subspace spanned by the first two principal components. The \emph{rain} and \emph{no rain} data points are represented as blue  circles and green squares, respectively.
\label{fig:subspace}}
\end{center}
\vspace{-1cm}
\end{figure}

We now analyze the weather- and time-related variables considered in this paper for the purpose of rainfall detection. Fig.~\ref{fig:subspace} shows the plot of samples against the first two principal components, similar to Figure \ref{fig:biplot}. Sample points are distinguished based on the rainfall data recorded by the weather station in order to provide us an intuition on the degree of separability of \emph{rain} and \emph{no-rain} data points. As is apparent from the plot, the lower dimensional subspace created via PCA does indeed facilitate discrimination between \emph{rain} and \emph{no-rain} observations. 

\section{Conclusions \& Future Work}
\label{sec:conc}
We have conducted a systematic analysis of various weather parameters measured in the tropics. We analyzed the correlation coefficients between $7$ different meteorological variables. We then used PCA and found that 4 principal components are needed to explain a cumulative variance of 85\%. This means that the individual variables are relatively uncorrelated and have their own significance in describing different weather phenomena. 

The subspace representation of data samples against PC1 and PC2 with the rain condition taken into consideration indicates the usefulness of  PCA in separating rain and no-rain cases; it can therefore be used in machine learning classification algorithms to detect rain. However, as discussed, the other principal components also have significant contributions to the total variance explained, and might be useful in further improving the classification results.

In the future, we plan to include sky images~\cite{IGARSS2015} and image-based features for detecting precipitation \cite{rainonset}. We also aim to propose a machine-learning based framework~\cite{Dev2016GRSM} to predict precipitation in a multi-modal manner, using weather- and image-based features.

\balance

\end{document}